\documentclass{elsart}
\usepackage{graphics}
\usepackage{graphicx}
\usepackage{epsfig}
\usepackage{amssymb}

\renewcommand{\theenumi}{(\arabic{enumi})}
\begin{document}

\begin{frontmatter}

\title{Lineshape of the $\Lambda(1405)$ Hyperon Measured Through its $\Sigma^0\pi^0$ Decay}

\author[Swierk]{I.\,Zychor},
\author[IKP]{M.\,B\"{u}scher},
\author[IKP]{M.\,Hartmann},
\author[Tbilisi,Erlangen]{A.\,Kacharava},
\author[IKP,Tbilisi]{I.\,Keshelashvili},
\author[Muenster]{A.\,Khoukaz},
\author[Bonn]{V.\,Kleber},
\author[Gatchina]{V.\,Koptev},
\author[Osaka]{Y.\,Maeda},
\author[Muenster]{T.\,Mersmann},
\author[Gatchina]{S.\,Mikirtychiants},
\author[IKP]{R.\,Schleichert},
\author[IKP]{H.\,Str\"oher},
\author[Gatchina]{Yu.\,Valdau},
\author[CW_College]{C.\,Wilkin\corauthref{cor1}}
\ead{cw@hep.ucl.ac.uk} \corauth[cor1]{Corresponding author.}
%
\address[Swierk]{The Andrzej So{\l}tan Institute for Nuclear Studies,
05-400 \'Swierk, Poland}
\address[IKP]{Institut f\"ur Kernphysik, Forschungszentrum J\"ulich,
  52425 J\"ulich, Germany}
\address[Tbilisi]{High Energy Physics Institute, Tbilisi State University,
0186 Tbilisi, Georgia}
\address[Erlangen]{Physikalisches Institut II, Universit{\"a}t
Erlangen-N{\"u}rnberg, 91058 Erlangen, Germany}%
\address[Muenster]{Institut f\"ur Kernphysik, Universit\"at
M\"unster, 48149 M\"unster, Germany}%
\address[Bonn]{Physikalisches Institut, Universit\"at
Bonn, 53115 Bonn, Germany}
\address[Gatchina]{High Energy Physics Department, Petersburg Nuclear
Physics Institute, 188350 Gatchina, Russia}
\address[Osaka]{Research Center for Nuclear Physics, Osaka
University, Ibaraki, Osaka 567-0047, Japan}
\address[CW_College]{Physics and Astronomy Department, UCL, London, WC1E 6BT, UK}
%
%
\begin{abstract}
The $pp \rightarrow  pK^+ Y^0$ reaction has been studied for
hyperon masses $m(Y^0) \leq 1540\,$MeV/c$^2$ at COSY-J\"ulich by
using a 3.65\,GeV/c circulating proton beam incident on an
internal hydrogen target. Final states comprising two protons, one
positively charged kaon and one negatively charged pion have been
identified with the ANKE spectrometer. Such configurations are
sensitive to the production of the ground state $\Lambda$ and
$\Sigma^0$ hyperons as well as the $\Sigma^0(1385)$ and
$\Lambda(1405)$ resonances. Applying invariant-- and missing--mass
techniques, the two overlapping excited states could be well
separated, though with limited statistics. The shape and
position of the $\Lambda(1405)$ distribution, reconstructed
cleanly in the $\Sigma^0\pi^0$ channel, are similar to those found
from other decay modes and there is no obvious mass shift. This
finding constitutes a challenging test for models that predict
$\Lambda(1405)$ to be a two-state resonance.
\end{abstract}

\end{frontmatter}

The excited states of the nucleon are a topical field of research,
since the full spectrum contains deep-rooted information about the
underlying strong colour force acting between the quarks and
gluons. In addition to searching for missing resonances predicted
by quark models~\cite{Capstick00}, it is important to understand
the structure of certain well established states, such as the
$\Lambda(1405)$ hyperon resonance.

Although a four--star resonance~\cite{PDG}, and known already for
many years, the dynamics of the $\Lambda(1405)$ are still not
fully understood. Within the quark model it can be explained as a
$P$--wave $q^3$ baryon~\cite{Isgur}. It is also widely discussed
as a candidate for a $\bar{K}N$ molecular state~\cite{DalitzTuan},
or for one with a more intrinsic $q^4 \bar{q}$ pentaquark
structure~\cite{Inoue}. If the $\Lambda(1405)$ is a dynamically
generated resonance produced \textit{via} $\bar{K}N$ rescattering
within a coupled--channel formalism~\cite{Nacher,Oller}, it may
consist of two overlapping $I=0$ states~\cite{Jido,Magas,Geng}.
Its decay spectrum would then depend upon the production reaction.
Due to the opening of the $\bar{K}N$ channels, the $\Lambda(1405)$
lineshape is not represented satisfactorily by a Breit--Wigner
resonance~\cite{DalitzTuan,Hem,Dalitz,Thomas}. Nevertheless, if
the $\Lambda(1405)$ were a single quantum state, as in the quark
model or molecular pictures, its lineshape should be independent
of the production method.

Part of the difficulty in elucidating the nature of the
$\Lambda(1405)$ is due to it overlapping the nearby
$\Sigma^0(1385)$. The interference between these two states can
distort significantly the $\Sigma^+\pi^-$ and $\Sigma^-\pi^+$
spectra~\cite{Nacher}, for which there are experimental
indications~\cite{Ahn}. This interference can be eliminated by
taking the average of $\Sigma^+\pi^-$ and $\Sigma^-\pi^+$
data~\cite{Hem} but the cleanest approach is through the measurement of
the $\Sigma^0 \pi^0$ channel, since isospin forbids this for
$\Sigma^0(1385)$ decay. This is the technique that we want to develop here
and, although our statistics are rather poor, these are already sufficient
to yield promising results.

We have used data obtained during high statistics
$\phi$--production measurements with the ANKE
spectrometer~\cite{PHI} to study the excitation and decay of
low--lying hyperon resonances in $pp$ collisions at a beam
momentum of 3.65~GeV/c in an internal--ring experiment at
COSY--J\"ulich. A dense hydrogen cluster--jet gas
target was used and over a four--week period this
yielded an integrated luminosity of $L=(69\pm 10)~\textrm{pb}^{-1}$,
as determined from elastic $pp$ scattering that was measured in
parallel and compared with the SAID 2004 solution~\cite{SAID}.

The detection systems of the magnetic three--dipole spectrometer
ANKE simultaneously register and identify both negatively and
positively charged particles~\cite{ANKE_NIM}. Forward (Fd) and
side--wall (Sd) counters were used for protons, telescopes and
side--wall scintillators for $K^+$, and scintillators for $\pi^-$.
Since the efficiencies of the detectors are constant to 2\%
($\sigma$) across the momentum range of registered particles, any
uncertainty in this can be neglected in the further analysis.

The basic principle of the experiment is the search for four--fold
coincidences, comprising two protons, one positively charged kaon
and one negatively charged pion, i.e., $pp \rightarrow p K^+ p
\pi^- X^0$. Such a configuration can correspond, e.g., to the
following reaction chains involving the $\Sigma^0(1385)$ and
$\Lambda(1405)$ as intermediate states: \\[1ex] %
(1)\ $pp \rightarrow p K^+ \Sigma^0(1385) \rightarrow p K^+
\Lambda \pi^0 \rightarrow p K^+ p \pi^- \pi^0$ \\ %
(2)\ $pp\rightarrow p K^+ \Lambda(1405) \rightarrow p K^+ \Sigma^0
\pi^0 \rightarrow p K^+ \Lambda \gamma \pi^0 \rightarrow p K^+ p
\pi^- \gamma \pi^0$.

In the $\Sigma^0(1385)$ case, the residue is $X^0 = \pi^0$, while
for the $\Lambda(1405)$, $X^0 = \pi^0 \gamma$. The resonances
overlap significantly because the widths of 36\,MeV/c$^2$  for
$\Sigma^0(1385)$ and 50\,MeV/c$^2$ for $\Lambda(1405)$ are much
larger than the mass difference~\cite{PDG}. The strategy to
discriminate between them is to: (\textit{i}) detect and identify
four charged particles $p_{Fd}$, $p_{Sd}$, $K^+$ and $\pi^-$ in
coincidence, thereby drastically reducing the accidental
background at the expense of statistics, (\textit{ii}) select
those events for which the mass of a ($p_{Sd} \pi^-$) pair
corresponds to that of the $\Lambda$, (\textit{iii}) select the
mass of the residue $m(X^0)$ to be that of the $\pi^0$ to tag the
$\Sigma^0(1385)$, and $m(X^0) > m(\pi^0)+55\,\textrm{MeV/c}^2$ for
the $\Lambda(1405)$.

\begin{center}
\begin{figure}[ht]
\psfig{file=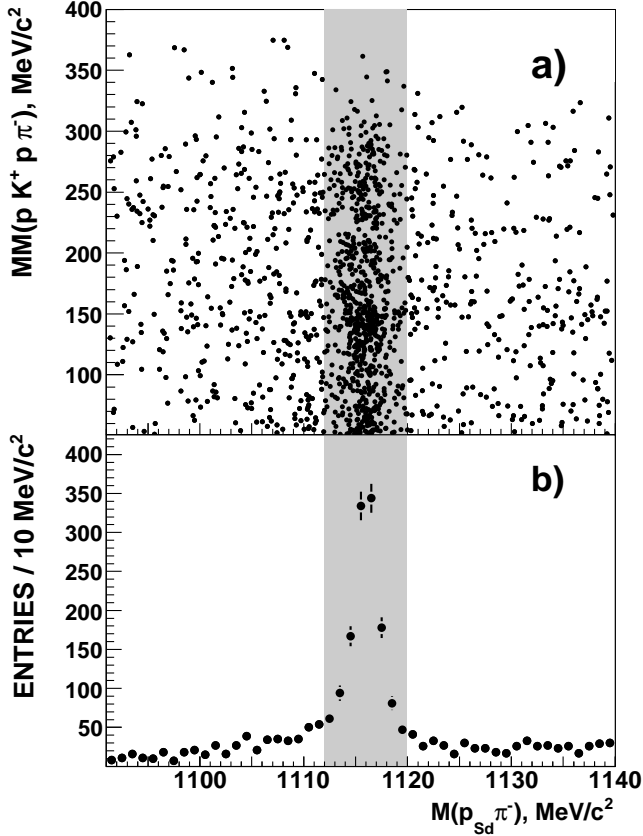,height=14cm}\\
\caption{a)  Missing mass $MM(pK^+p\pi^-)$ \textit{versus}
invariant mass $M(p_{Sd}\pi^-)$. The shaded vertical box shows the
band used to select the $\Lambda$. b) The projection of all the
events from panel a) onto the $M(p_{Sd}\pi^-)$ axis shows a clear
$\Lambda$ peak with a FWHM projection of $\sim 5\,$MeV/c$^2$ and a
slowly varying background. } \label{fig1}
\end{figure}
\end{center}

Figure~\ref{fig1}a shows the two--dimensional distribution of
the four--particle missing mass $MM(p K^+\pi^-p)$ of the $p_{Sd}
\pi^-$ pairs \textit{versus} the invariant mass $M(p_{Sd}\pi^-)$.
A vertical band corresponding to the $\Lambda$, is visible around
a mass of 1116\,MeV/c$^2$. The features of this band are
illustrated clearly in the projection onto the $M(p_{Sd}\pi^-)$
axis shown in Fig.~\ref{fig1}b. The $\Lambda$ peak, with a
FWHM of $\sim 5\,$MeV/c$^2$, sits on a slowly varying background,
much of which arises from a false $p\pi^-$ association (the
combinatorial background).

\begin{figure}[ht]
\psfig{file=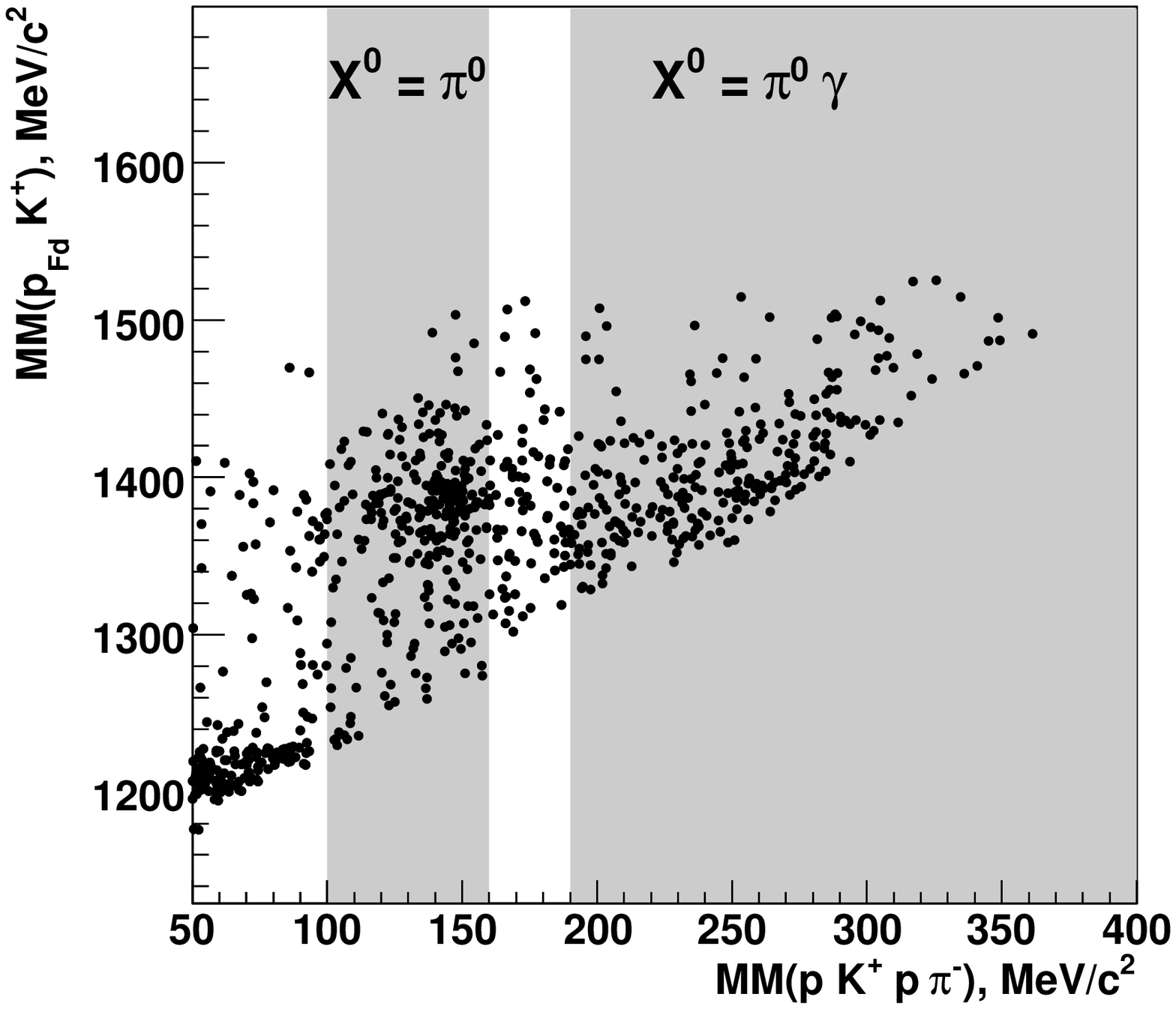,height=8cm}\\
\caption{Missing mass $MM(p_{Fd}K^+)$ \textit{versus}
$MM(pK^+p\pi^-)$. A clear concentration of $\pi^0$ events is seen,
though with a central value of the mass $\sim 8\,$MeV/c$^2$ too
high, a deviation that is consistent with the resolution expected
for a four--particle missing mass. The left shaded vertical box
covers this $\pi^0$ region and the right one has $MM(p K^+p\pi^-)
> 190\,\textrm{MeV}/c^2$ originating, e.g., from $\pi^0\gamma$ and
$\pi\pi$. } \label{fig2}
\end{figure}

Data within the invariant--mass window 1112--1120\,MeV/c$^2$ were
retained for further analysis and, in Fig.~\ref{fig2},
$MM(p_{Fd}K^+)$ is plotted against $MM(p K^+p\pi^-)$ for these
events. The triangular--shaped domain arises from the constraint
$MM(p_{Fd}K^+)\geq MM(p K^+p\pi^-)+m(\Lambda)$. Despite the lower
limit of 50\,MeV/c$^2$ on $MM(p K^+p\pi^-)$, there is a background
from $\Sigma^0$ production at the bottom of the triangle, but this
can be easily cut away. The enhancement for $MM(p_{Fd}K^+)\sim
1400\,$MeV/c$^2$ corresponds to $\Sigma^0(1385)$ and
$\Lambda(1405)$ production. The two vertical bands show the
four--particle missing--mass $MM(p K^+p\pi^-)$ criteria used to
separate the $\Sigma^0(1385)$ from the $\Lambda(1405)$. The left
band is optimised to identify a $\pi^0$ whereas, in view of the
missing--mass resolution, the right one selects masses
significantly greater than $m(\pi^0)$.

\begin{figure}[ht]
\psfig{file=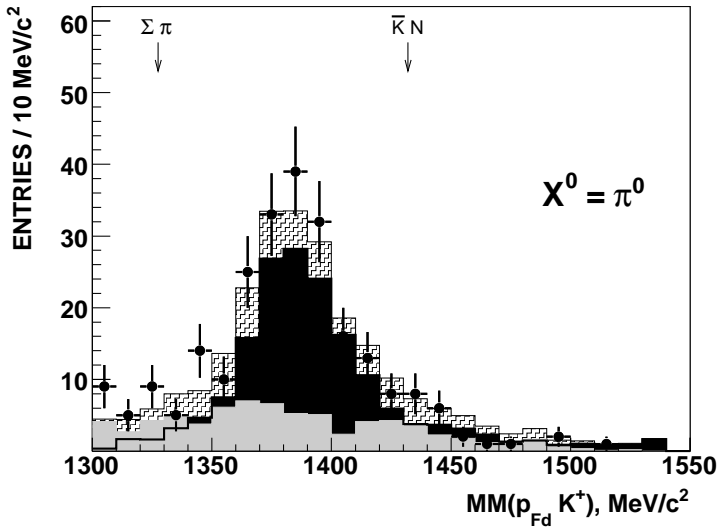,height=8cm} \caption{ Missing--mass
$MM(p_{Fd}K^+)$ distribution for the $pp\rightarrow pK^+ p \pi^-
X^0$ reaction for events with $M(p_{Sd}\pi^-)\approx m(\Lambda)$
and for $MM(p K^+p\pi^-)\approx m(\pi^0)$. Experimental points
with statistical errors are compared to the shaded histogram of
the fitted overall Monte Carlo simulations. The simulation
includes resonant contributions (solid--black) and non--resonant
phase--space production (solid--grey). The structure in the latter
arises from the various channels considered. Arrows indicate the $\Sigma
\pi$ and $\bar{K} N$ thresholds. } \label{fig3}
\end{figure}

Since the properties of the $\Sigma^0(1385)$ are
undisputed~\cite{PDG}, we first present and discuss results for
this hyperon as a test case for the $\Lambda(1405)$ analysis. In
Fig.~\ref{fig3} we show the experimental missing--mass $MM(p_{Fd}
K^+)$ spectrum for events within the $\pi^0$--band of
Fig.~\ref{fig2}. When this is fit with a Breit--Wigner
distribution plus a linear background, a mass of $M=
(1384\pm10)\,\textrm{MeV/c}^2$ and a width of $\Gamma\sim
40\,$MeV/c$^2$ are obtained, in good agreement with the PDG
values~\cite{PDG}. The resonance is located half way between the
$\Sigma \pi$ and $\bar{K} N$ thresholds, indicated by arrows in
Fig.~\ref{fig3}, and no significant influence of either threshold
is observed in the data.

To investigate possible contributions to the spectrum other than
from the $\Sigma^0(1385)$ excitation, Monte Carlo simulations were
performed for backgrounds from non--resonant and resonant
production. The first group of reactions includes processes such
as $pp \rightarrow NK^+\pi X (\gamma)$ and $pp \rightarrow
NK^+\pi\pi X (\gamma)$, with $X$ representing any allowed
$\Lambda$ or $\Sigma$ hyperon. The second group comprises
$\Lambda(1405)$ and $\Lambda(1520)$ hyperon production. The
simulations, based on the GEANT3 package, were performed in a
similar manner to those in Ref.~\cite{Y1480}. Events were
generated according to phase space using relativistic
Breit--Wigner parametrisations for the known hyperon
resonances~\cite{PDG}. Their relative contributions were deduced
by fitting the experimental data, giving the results shown by the
histograms of Fig.~\ref{fig3}. Also included is a small contribution
from the $\Lambda(1405)$ channel, arising from the tail of the
missing--mass events in Fig.~\ref{fig2} leaking into the $\pi^0$
region. As expected, the $\Sigma^0(1385)$ peak dominates over a
small and smooth background.

In order to estimate the total $\Sigma^0(1385)$ production cross
section we used the overall detector efficiency of $\sim55\%$ and
the cumulative branching ratio of $56\%$ for the $\Sigma^0(1385)$
decay chain corresponding to reaction~(1). With the calculated
acceptance of $\sim 2\times10^{-6}$ and the number of
$\Sigma^0(1385)$ events equal to $170 \pm 26$, we find
\[\sigma_{\rm{tot}}(pp \rightarrow p K^+ \Sigma^0(1385)) = (4.0
\pm 1.0_{\rm{stat}}\pm 1.6_{\rm{syst}})\,\mu\textrm{b}.\]%
at $p_{\rm{beam}}=3.65\,$GeV/c. The systematic uncertainty in the
fitting procedure and cross section evaluation was estimated by
varying some of the event selection parameters, such as the width
of the $MM(p K^+p\pi^-)$ bands or the range for the $\Lambda$ peak
(see Fig.~\ref{fig1}), or the non--resonant background in
Fig.~\ref{fig3}. The cross section is only a little lower
than at 6\,GeV/c, $(7 \pm 1)\,\mu$b~\cite{Klein}, whereas that for
$pp\to pK^+\Lambda$ increases by a factor of four over a similar
change in excess energy~\cite{Sib}.

\begin{figure}[t]
\psfig{file=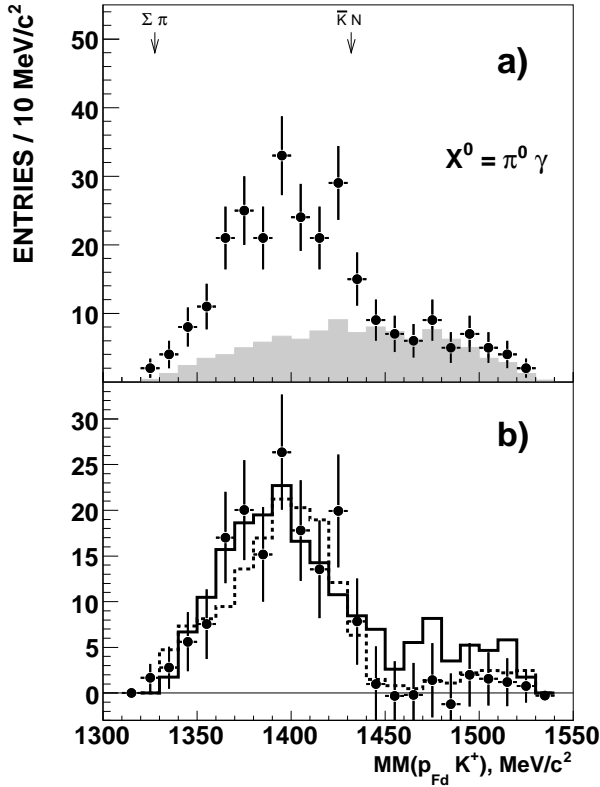,width=9.3cm}%
\caption{a) Missing--mass $MM(p_{Fd}K^+)$ distribution for the $pp
\rightarrow pK^+p \pi^- X^0$ reaction for events with
$M(p_{Sd}\pi^-) \approx m(\Lambda)$ and $MM(p K^+p\pi^-) >
190\,\textrm{MeV/c}^2$. Experimental points with statistical
errors are compared to the shaded histogram of the fitted
non--resonant Monte Carlo simulation. b) The
background--subtracted lineshape of the $\Lambda(1405)$ decaying
into $\Sigma^0 \pi^0$ (points) compared to $\pi^- p \rightarrow
K^0 (\Sigma \pi)^0$~\cite{Thomas} (solid line) and
$K^-p\rightarrow \pi^+\pi^-\Sigma^+\pi^-$~\cite{Hem} (dotted line)
data.} \label{fig4}
\end{figure}

Turning now to the $\Lambda(1405)$, simulations show that the
$\Sigma^0(1385)$ does not contaminate the missing--mass $MM(p
K^+p\pi^-)$ range above 190\,MeV/c$^2$. This point is crucial
since it allows us to obtain a clean separation of the
$\Sigma^0(1385)$ and $\Lambda(1405)$. There is the possibility of
some contamination from the $pK^+\Lambda(\pi\pi)^0$ channel but
there is only a limited amount of the five--body phase space
available near the maximum missing mass. Simulations also show
that the ANKE acceptance varies only marginally in the mass range
around 1400\,MeV/c$^2$. The corresponding experimental
missing--mass $MM(p_{Fd}K^+)$ spectrum is shown in
Fig.~\ref{fig4}a. The asymmetric distribution, which peaks around
1400\,MeV/c$^2$, has a long tail on the high missing--mass side
that extends up to the kinematical limit.

In order to extract the $\Lambda(1405)$ distribution from the
measured $\Sigma^0 \pi^0$ decay, a different strategy has
been applied, where we first fit the non--resonant
contributions to the experimental data. The fit was performed
for $1440 <  MM(p_{Fd}K^+) <
1490\,$MeV/c$^2$ to exclude heavier hyperon resonances, such
as the $\Lambda(1520)$. The resulting non--resonant background is
indicated by the shaded histogram in Fig.~\ref{fig4}a. When
this is subtracted from the data we obtain the distribution
shown as experimental points in Fig.~\ref{fig4}b.

Our background--subtracted data exhibit a prominent structure
around 1400\,MeV/c$^2$. There is no indication of a second near
1500\,MeV/c$^2$, which might result from the production of the
$\Lambda(1520)$~\cite{Hem}. The excess of at most 20~events for
$MM(p_{Fd}K^+)>1490\,\textrm{MeV/c}^2$ leads to an upper limit for
the $\Lambda(1520)$ production cross section of $\sigma_{\rm{tot}}
<~0.2\,\mu$b. The smallness of the signal in this case would be
largely due to the low branching of only 9\% into this channel.
There is no evidence either for a significant contribution from
the $Y^{0*}(1480)$ hyperon~\cite{Y1480}. If this state were the
same as the one--star $\Sigma^0(1480)$ of Ref.\cite{PDG}, the
decay into $\Sigma^0 \pi^0$ would be forbidden. However, this
state is also not seen in the $K^- p \rightarrow \pi^0 \pi^0
\Lambda$ reaction~\cite{Prakhov_1480}.

We finally turn to the contribution from lower missing masses.
From the number of events with
$1320<MM(p_{Fd}K^+)<1440\,$MeV/c$^2$, equal to $156 \pm 23$, we find
a total production cross section of
\[\sigma_{\rm{tot}}(pp \rightarrow p K^+ \Lambda(1405)) = (4.5
\pm 0.9_{\rm{stat}}\pm 1.8_{\rm{syst}})\,\mu\textrm{b}\]
at $p_{\rm{beam}}=3.65\,$GeV/c. The cumulative branching ratio for
the $\Lambda(1405)$ decay chain of reaction~(2) of $21\%$ and the
acceptance of $\sim4\times10^{-6}$ have been included, as well as
the overall detection efficiency of $\sim55\%$.

The $(\Sigma \pi)^0$ invariant--mass distributions have been
studied in two hydrogen bubble chamber experiments. Thomas
\textit{et al.}~\cite{Thomas} found $\sim 400$
$\Sigma^+\pi^-$ or $\Sigma^-\pi^+$ events corresponding to the $\pi^- p
\rightarrow K^0 \Lambda(1405)\rightarrow K^0 (\Sigma \pi)^0$
reaction at a beam momentum of 1.69\,GeV/c. Hemingway~\cite{Hem}
used a 4.2\,GeV/c kaon beam to investigate $K^- p \rightarrow
\Sigma^+(1660) \pi^- \rightarrow \Lambda(1405) \pi^+ \pi^-
\rightarrow  (\Sigma ^\pm \pi^\mp) \pi^+ \pi^-$. For the
$\Sigma^-\pi^-\pi^+\pi^+$ final state, the $\Sigma^- \pi^+$ mass
spectrum is distorted by the confusion between the two positive
pions. Thus, in the comparison with our data, we use only the
$\Sigma^+ \pi^-$ distribution, which contains 1106
events~\cite{Hem}.

In Fig.~\ref{fig4}b our experimental points are compared to the
results of Thomas and Hemingway, which have been normalised by
scaling their values down by factors of $\sim$3 and $\sim$7,
respectively. The effect of the $\bar{K} N$ threshold is apparent
in these published data, with the $\Lambda(1405)$ mass
distribution being distorted by the opening of this channel.
Despite the very different production mechanisms, the three
distributions have consistent shapes. A fit of one to either of
the others leads to a $\chi^2$/\textit{ndf} of the order of unity
though, as pointed out in Ref.~\cite{Nacher}, for $\Sigma^+\pi^-$
production~\cite{Hem} there is likely to be some residual
distortion from $I=1$ channels. The $K^- p \rightarrow
\Lambda(1405) \pi^0 \rightarrow \Sigma^0 \pi^0 \pi^0$ data yield a
somewhat different distribution~\cite{Prakhov_1405} but, as noted
in this reference, the uncertainty as to which $\pi^0$ originated
from the $\Lambda(1405)$ \textit{``smears the resonance signal in
the spectra''}. The situation is therefore very similar to that of
the Hemingway $\Sigma^-\pi^-\pi^+\pi^+$ data~\cite{Hem} and such
results can only be interpreted within the context of a specific
reaction model, such as that of Ref.~\cite{Magas}.

Models based on unitary chiral perturbation theory find two poles
in the neighborhood of the $\Lambda(1405)$ which evolve from a
singlet and an octet in the exact SU(3) limit~\cite{Jido,Magas}.
One has a mass of 1390\,MeV/c$^2$ and a width of 130\,MeV/c$^2$
and couples preferentially to $\Sigma \pi$. The narrower one,
located at 1425\,MeV/c$^2$, couples more strongly to $\bar{K} N$,
whose threshold lies at $\sim$~1432\,MeV/c$^2$. Both states may
contribute to the experimental distributions, and it is their
relative population, which depends upon the production mechanism,
that will determine the observed lineshape. Our experimental
findings show that the properties (mass, width, and shape) of the
$\Lambda(1405)$ resonance are essentially identical for these
three different production modes.

In summary, we have measured the excitation of the
$\Sigma^0(1385)$ and $\Lambda(1405)$ hyperon resonances in
proton--proton collisions at a beam momentum of 3.65\,GeV/c. We
have succeeded in unambiguously separating the two states through
their $\Lambda \pi^0$ and $\Sigma^0 \pi^0 \rightarrow \Lambda
\gamma \pi^0$ decay modes. Cross sections of the order of a few
$\mu$b have been deduced for both resonances. The $\Lambda(1405)$,
as measured through its $\Sigma^0\pi^0$ decay, has a shape that is
consistent with data on the charged decays~\cite{Hem,Thomas}, with
a mass of $\sim 1400\,$MeV/c$^2$ and width of $\sim
60\,$MeV/c$^2$. This might suggest that, if there are two states
present in this region, then the reaction mechanisms in the three
cases are preferentially populating the same one. However, by
identifying particular reaction mechanisms, proponents of the
two--state solution can describe the shape of the distribution
that we have found~\cite{Geng}.

The $\Sigma^0\pi^0$ channel is by far the cleanest for the
observation of the $\Lambda(1405)$ since it is not contaminated by
the $\Sigma(1385)$ nor the confusion regarding the identification
of the pion from its decay. However, although we have shown that
the method works in practice, in view of our limited
statistics, further data are clearly needed. The decay
$\Lambda(1405)\to \Sigma^0 \pi^0 \rightarrow \Lambda \gamma \pi^0$
can be detected directly in electromagnetic calorimeters.
Corresponding measurements are under way in $\gamma p$~reactions
(CB/TAPS~at~ELSA~\cite{ELSA}, SPring$-$8/LEPS~\cite{LEPS}) and are
also planned in $pp$ interactions with WASA~at~COSY~\cite{WASA}.

We acknowledge many very useful discussions with E.~Oset. We also
thank all other members of the ANKE collaboration and the COSY
accelerator staff for their help during the data taking. This work
has been supported by COSY-FFE Grant, BMBF, DFG and Russian
Academy of Sciences.

\end{document}